\documentstyle[epsfig]{elsart}
\newcommand{\be}{\begin{equation}}
\newcommand{\ee}{\end{equation}}
\newcommand{\bea}{\begin{eqnarray}}
\newcommand{\eea}{\end{eqnarray}}
\newcommand{\bean}{\begin{eqnarray*}}
\newcommand{\eean}{\end{eqnarray*}}

\newcommand{\cD} {{\cal D}}
\newcommand{\cL} {{\cal L}}
\newcommand{\Tr} {{\mbox{Tr~}}}
\newcommand{\RE} {{\mbox{Re~}}}

\newcommand{\pre} {\mbox{\em Phys. Rep.\/ }}
\newcommand{\np}  {\mbox{\em Nucl. Phys.\/ }}
\newcommand{\pr}  {\mbox{\em Phys. Rev.\/ }}
\newcommand{\prl} {\mbox{\em Phys. Rev. Lett.\/ }}
\newcommand{\pl}  {\mbox{\em Phys. Lett.\/ }}
\newcommand{\ap}  {\mbox{\em Ann. Phys. \/ }}
\newcommand{\cmp} {\mbox{\em Comm. Math. Phys.\/ }}
\newcommand{\rmp} {\mbox{\em Rev. Mod. Phys.\/ }}
\newcommand{\jmp} {\mbox{\em J. Mod. Phys.\/ }}

\newcommand{\plaq}{\setlength{\unitlength}{.5cm}\raisebox{-.2cm}{
   \begin{picture}(1.2,1.2)(-.6,-.6)
   \basispl\basisar
   \put(-.5,-.5){\circle*{.2}}
   \put(-.55,-.55){\makebox(0,0)[tr]{\footnotesize $x$}}
   \put(-.55,0){\makebox(0,0)[r]{\footnotesize $\nu$}}
   \put(0,-.55){\makebox(0,0)[t]{\footnotesize $\mu$}}
   \end{picture}}}

\newcommand{\basispl}{
   \put(-.5,-.5){\line(1,0){1}}
   \put(.5,-.5){\line(0,1){1}}
   \put(.5,.5){\line(-1,0){1}}
   \put(-.5,.5){\line(0,-1){1}} }

\newcommand{\basisar}{
   \put(0,-.5){\vector(1,0){0}}
   \put(.5,0){\vector(0,1){0}}
   \put(0,.5){\vector(-1,0){0}}
   \put(-.5,0){\vector(0,-1){0}} }

\newcommand{\twooneplaq}{\setlength{\unitlength}{.5cm}
   \raisebox{-.2cm}{
   \begin{picture}(2.2,1.2)(-1.1,-.6)
   \put(-1,-.5){\line(1,0){2}}
   \put(-1,.5){\line(1,0){2}}
   \put(-1,-.5){\line(0,1){1}}
   \put(1,-.5){\line(0,1){1}}
   \put(-0.5,-.5){\vector(1,0){0}}
   \put(0.5,-.5){\vector(1,0){0}}
   \put(-0.5,.5){\vector(-1,0){0}}
   \put(0.5,.5){\vector(-1,0){0}}
   \put(-1,.0){\vector(0,-1){0}}
   \put(1,.0){\vector(0,1){0}}
   \multiput(-1,-.5)(1,0){3}{\circle*{.2}}
   \multiput(-1,.5)(1,0){3}{\circle*{.2}}
   \put(-1.0,-.55){\makebox(0,0)[tr]{\footnotesize $x$}}
   \put(-1.0,0){\makebox(0,0)[r]{\footnotesize $\nu$}}
   \put(0,-.55){\makebox(0,0)[t]{\footnotesize $\mu$}}
   \end{picture}}}

\newcommand{\onetwoplaq}{\setlength{\unitlength}{1cm}\raisebox{-.5cm}{
   \begin{picture}(1.2,1.2)(-.6,-.6)
   \put(.25,-.5){\line(0,1){1}}
   \put(.25,.5){\line(-1,0){.5}}
   \put(-.25,.5){\line(0,-1){1}}
   \put(.2,0){\line(1,0){.05}}
   \put(-.25,-.5){\circle*{.1}}
   \put(-.25,.5){\circle*{.1}}
   \put(.25,-.5){\circle*{.1}}
   \put(.25,.5){\circle*{.1}}
   \put(-.25,0){\circle*{.1}}
   \put(.25,0){\circle*{.1}}
   \put(-.25,-.35){\vector(0,-1){0}}
   \put(-.25,.15){\vector(0,-1){0}}
   \put(-.1,.5){\vector(-1,0){0}}
   \put(.25,.35){\vector(0,1){0}}
   \put(.25,-.15){\vector(0,1){0}}
   \put(-.25,-.5){\line(1,0){.5}}
   \put(.1,-.5){\vector(1,0){0}}
   \put(-.3,-.55){\makebox(0,0)[tr]{\footnotesize $x$}}
   \put(-.3,0){\makebox(0,0)[r]{\footnotesize $\nu$}}
   \put(0,.8){\makebox(0,0)[t]{\footnotesize $\mu$}}
   \end{picture}}}

\newcommand{\twoplaq}{\setlength{\unitlength}{1cm}\raisebox{-.5cm}{
   \begin{picture}(1.2,1.2)(-.6,-.6)
   \basispl
   \put(-.5,-.5){\circle*{.1}}
   \put(-.5,.5){\circle*{.1}}
   \put(.5,-.5){\circle*{.1}}
   \put(.5,.5){\circle*{.1}}
   \put(0,-.5){\circle*{.1}}
   \put(0,.5){\circle*{.1}}
   \put(.5,0){\circle*{.1}}
   \put(-.5,0){\circle*{.1}}
   \put(-.25,-.5){\vector(1,0){0}}
   \put(.25,-.5){\vector(1,0){0}}
   \put(.5,-.25){\vector(0,1){0}}
   \put(.5,.25){\vector(0,1){0}}
   \put(-.25,.5){\vector(-1,0){0}}
   \put(.25,.5){\vector(-1,0){0}}
   \put(-.5,-.25){\vector(0,-1){0}}
   \put(-.5,.25){\vector(0,-1){0}}
   \put(-.55,-.55){\makebox(0,0)[tr]{\footnotesize $x$}}
   \put(-.55,0){\makebox(0,0)[r]{\footnotesize $\nu$}}
   \put(0,-.55){\makebox(0,0)[t]{\footnotesize $\mu$}}
   \end{picture}}}

\newcommand{\stapup}{\setlength{\unitlength}{.5cm}\raisebox{-.2cm}{
   \begin{picture}(1.2,1.2)(-.6,-.6)
   \put(.5,-.5){\line(0,1){1}}
   \put(.5,.5){\line(-1,0){1}}
   \put(-.5,.5){\line(0,-1){1}}
   \put(.5,0){\vector(0,-1){0}}
   \put(0,.5){\vector(1,0){0}}
   \put(-.5,0){\vector(0,1){0}}
   \put(-.5,-.5){\circle*{.2}}
   \put(-.55,-.55){\makebox(0,0)[tr]{\footnotesize $x$}}
   \put(-.55,0){\makebox(0,0)[r]{\footnotesize $\nu$}}
   \put(0,.55){\makebox(0,0)[b]{\footnotesize $\mu$}}
   \end{picture}}}

\newcommand{\stapdw}{\setlength{\unitlength}{.5cm}\raisebox{-.2cm}{
   \begin{picture}(1.2,1.2)(-.6,-.6)
   \put(.5,-.5){\line(0,1){1}}
   \put(.5,-.5){\line(-1,0){1}}
   \put(-.5,.5){\line(0,-1){1}}
   \put(.5,0){\vector(0,1){0}}
   \put(0,-.5){\vector(1,0){0}}
   \put(-.5,0){\vector(0,-1){0}}
   \put(-.5,.5){\circle*{.2}}
   \put(-.55,.75){\makebox(0,0)[tr]{\footnotesize $x$}}
   \put(-.55,0){\makebox(0,0)[r]{\footnotesize $\nu$}}
   \put(0,-.55){\makebox(0,0)[t]{\footnotesize $\mu$}}
   \end{picture}}}

\newcommand{\wplaqone}{\setlength{\unitlength}{1cm}\raisebox{-.5cm}{
   \begin{picture}(1.2,1.2)(-.6,-.6)
   \put(.25,-.5){\line(0,1){1}}
   \put(.25,.5){\line(-1,0){.5}}
   \put(-.25,.5){\line(0,-1){1}}
   \put(-.25,0){\line(1,0){.1}}
   \put(-.10,0){\line(1,0){.1}}
   \put(.05,0){\line(1,0){.1}}
   \put(.2,0){\line(1,0){.05}}
   \put(-.25,-.5){\circle*{.1}}
   \put(-.25,.5){\circle*{.1}}
   \put(.25,-.5){\circle*{.1}}
   \put(.25,.5){\circle*{.1}}
   \put(-.25,0){\circle*{.1}}
   \put(.25,0){\circle*{.1}}
   \put(-.25,-.15){\vector(0,1){0}}
   \put(-.25,.35){\vector(0,1){0}}
   \put(.1,.5){\vector(1,0){0}}
   \put(.25,.15){\vector(0,-1){0}}
   \put(.25,-.35){\vector(0,-1){0}}
   \put(-.3,-.55){\makebox(0,0)[tr]{\footnotesize $x$}}
   \put(-.3,0){\makebox(0,0)[r]{\footnotesize $\nu$}}
   \put(0,.8){\makebox(0,0)[t]{\footnotesize $\mu$}}
   \end{picture}}}

\newcommand{\wplaqtwo}{\setlength{\unitlength}{1cm}\raisebox{-.5cm}{
   \begin{picture}(1.2,1.2)(-.6,-.6)
   \put(-.5,-.25){\line(0,1){.5}}
   \put(-.5,.25){\line(1,0){1}}
   \put(.5,.25){\line(0,-1){.5}}
   \put(.5,-.25){\line(-1,0){.5}}
   \put(0,-.25){\line(0,1){.1}}
   \put(0,-.10){\line(0,1){.1}}
   \put(0,.05){\line(0,1){.1}}
   \put(0,.20){\line(0,1){.1}}
   \put(-.5,-.25){\circle*{.1}}
   \put(0,-.25){\circle*{.1}}
   \put(.5,-.25){\circle*{.1}}
   \put(-.5,.25){\circle*{.1}}
   \put(0,.25){\circle*{.1}}
   \put(-.5,.25){\circle*{.1}}
   \put(-.5,.1){\vector(0,1){0}}
   \put(-.15,.25){\vector(1,0){0}}
   \put(.35,.25){\vector(1,0){0}}
   \put(.5,-.1){\vector(0,-1){0}}
   \put(.2,-.25){\vector(-1,0){0}}
   \put(-.55,-.3){\makebox(0,0)[tr]{\footnotesize $x$}}
   \put(-.55,0){\makebox(0,0)[r]{\footnotesize $\nu$}}
   \put(-.25,.5){\makebox(0,0)[t]{\footnotesize $\mu$}}
   \end{picture}}}

\newcommand{\wplaqthree}{\setlength{\unitlength}{1cm}\raisebox{-.5cm}{
   \begin{picture}(1.2,1.2)(-.6,-.6)
   \put(-.5,-.25){\line(0,1){.5}}
   \put(-.5,.25){\line(1,0){1}}
   \put(.5,.25){\line(0,-1){.5}}
   \put(0,-.25){\line(-1,0){.5}}
   \put(0,-.25){\line(0,1){.1}}
   \put(0,-.10){\line(0,1){.1}}
   \put(0,.05){\line(0,1){.1}}
   \put(0,.20){\line(0,1){.1}}
   \put(-.5,-.25){\circle*{.1}}
   \put(0,-.25){\circle*{.1}}
   \put(.5,-.25){\circle*{.1}}
   \put(-.5,.25){\circle*{.1}}
   \put(0,.25){\circle*{.1}}
   \put(-.5,.25){\circle*{.1}}
   \put(-.5,.1){\vector(0,1){0}}
   \put(.35,.25){\vector(1,0){0}}
   \put(-.15,.25){\vector(1,0){0}}
   \put(.5,-.05){\vector(0,-1){0}}
   \put(-.35,-.25){\vector(-1,0){0}}
   \put(.1,-.35){\makebox(0,0)[tr]{\footnotesize $x$}}
   \put(-.55,0){\makebox(0,0)[r]{\footnotesize $\nu$}}
   \put(.37,.42){\makebox(0,0)[r]{\footnotesize $\mu$}}
   \end{picture}}}

\newcommand{\wplaqfour}{\setlength{\unitlength}{1cm}\raisebox{-.5cm}{
   \begin{picture}(1.2,1.2)(-.6,-.6)
   \put(.25,-.5){\line(0,1){1}}
   \put(.25,-.5){\line(-1,0){.5}}
   \put(-.25,.5){\line(0,-1){1}}
   \put(-.25,0){\line(1,0){.1}}
   \put(-.10,0){\line(1,0){.1}}
   \put(.05,0){\line(1,0){.1}}
   \put(.2,0){\line(1,0){.05}}
   \put(-.25,-.5){\circle*{.1}}
   \put(-.25,.5){\circle*{.1}}
   \put(.25,-.5){\circle*{.1}}
   \put(.25,.5){\circle*{.1}}
   \put(-.25,0){\circle*{.1}}
   \put(.25,0){\circle*{.1}}
   \put(-.25,-.35){\vector(0,-1){0}}
   \put(-.25,.15){\vector(0,-1){0}}
   \put( .1,-.5){\vector(1,0){0}}
   \put(.25,.35){\vector(0,1){0}}
   \put(.25,-.15){\vector(0,1){0}}
   \put(-.3,.75){\makebox(0,0)[tr]{\footnotesize $x$}}
   \put(-.3,0){\makebox(0,0)[r]{\footnotesize $\nu$}}
   \put(0,-.55){\makebox(0,0)[t]{\footnotesize $\mu$}}
   \end{picture}}}

\newcommand{\wplaqfive}{\setlength{\unitlength}{1cm}\raisebox{-.5cm}{
   \begin{picture}(1.2,1.2)(-.6,-.6)
   \put(-.5,-.25){\line(0,1){.5}}
   \put(0.0,.25){\line(1,0){.5}}
   \put(-.5,-.25){\line(1,0){1}}
   \put(.5,.25){\line(0,-1){.5}}
   \put(.5,-.25){\line(-1,0){.5}}
   \put(0,-.25){\line(0,1){.1}}
   \put(0,-.10){\line(0,1){.1}}
   \put(0,.05){\line(0,1){.1}}
   \put(0,.20){\line(0,1){.1}}
   \put(-.5,-.25){\circle*{.1}}
   \put(0,-.25){\circle*{.1}}
   \put(.5,-.25){\circle*{.1}}
   \put(-.5,.25){\circle*{.1}}
   \put(0,.25){\circle*{.1}}
   \put(-.5,.25){\circle*{.1}}
   \put(-.50,-.10){\vector(0,-1){0}}
   \put(-.15,-.25){\vector(1,0){0}}
   \put( .35,-.25){\vector(1,0){0}}
   \put( .50, .00){\vector(0,1){0}}
   \put( .25, .25){\vector(-1,0){0}}
   \put(-.55,.5){\makebox(0,0)[tr]{\footnotesize $x$}}
   \put(-.55,0){\makebox(0,0)[r]{\footnotesize $\nu$}}
   \put( .25,.5){\makebox(0,0)[t]{\footnotesize $\mu$}}
   \end{picture}}}

\newcommand{\wplaqsix}{\setlength{\unitlength}{1cm}\raisebox{-.5cm}{
   \begin{picture}(1.2,1.2)(-.6,-.6)
   \put(-.5,-.25){\line(0,1){.5}}
   \put(-.5,-.25){\line(1,0){1}}
   \put(.5,.25){\line(0,-1){.5}}
   \put(0,.25){\line(-1,0){.5}}
   \put(0,-.25){\line(0,1){.1}}
   \put(0,-.10){\line(0,1){.1}}
   \put(0,.05){\line(0,1){.1}}
   \put(0,.20){\line(0,1){.1}}
   \put(-.5,-.25){\circle*{.1}}
   \put(0,-.25){\circle*{.1}}
   \put(.5,-.25){\circle*{.1}}
   \put(-.5,.25){\circle*{.1}}
   \put(0,.25){\circle*{.1}}
   \put(-.5,.25){\circle*{.1}}
   \put(-.15, .25){\vector(-1,0){0}}
   \put(-.50,-.10){\vector(0,-1){0}}
   \put(-.15,-.25){\vector(1,0){0}}
   \put( .35,-.25){\vector(1,0){0}}
   \put( .50, .10){\vector(0,1){0}}
   \put(.1,.5){\makebox(0,0)[tr]{\footnotesize $x$}}
   \put(-.55,0){\makebox(0,0)[r]{\footnotesize $\nu$}}
   \put(-.30,.42){\makebox(0,0)[r]{\footnotesize $\mu$}}
   \end{picture}}}

\newcommand{\twoone}{\setlength{\unitlength}{1cm}\raisebox{-.5cm}{
   \begin{picture}(1.2,1.2)(-.6,-.6)
   \put(0,-.5){\line(1,0){.5}}
   \put(.5,-.5){\line(0,1){1}}
   \put(.5,.5){\line(-1,0){1}}
   \put(-.5,.5){\line(0,-1){1}}
   \put(-.5,-.5){\circle*{.1}}
   \put(-.5,.5){\circle*{.1}}
   \put(.5,-.5){\circle*{.1}}
   \put(.5,.5){\circle*{.1}}
   \put(0,-.5){\circle*{.1}}
   \put(0,.5){\circle*{.1}}
   \put(.5,0){\circle*{.1}}
   \put(-.5,0){\circle*{.1}}
   \put(.25,-.5){\vector(-1,0){0}}
   \put(.5,-.25){\vector(0,-1){0}}
   \put(.5,.25){\vector(0,-1){0}}
   \put(-.25,.5){\vector(1,0){0}}
   \put(.25,.5){\vector(1,0){0}}
   \put(-.5,-.25){\vector(0,1){0}}
   \put(-.5,.25){\vector(0,1){0}}
   \put(-.55,-.55){\makebox(0,0)[tr]{\footnotesize $x$}}
   \put(-.55,0){\makebox(0,0)[r]{\footnotesize $\nu$}}
   \put(.2,-.55){\makebox(0,0)[t]{\footnotesize $\mu$}}
   \end{picture}}}

\newcommand{\twotwo}{\setlength{\unitlength}{1cm}\raisebox{-.5cm}{
   \begin{picture}(1.2,1.2)(-.6,-.6)
   \put(-.5,-.5){\line(1,0){.5}}
   \put(.5,-.5){\line(0,1){1}}
   \put(.5,.5){\line(-1,0){1}}
   \put(-.5,.5){\line(0,-1){1}}
   \put(-.5,-.5){\circle*{.1}}
   \put(-.5,.5){\circle*{.1}}
   \put(.5,-.5){\circle*{.1}}
   \put(.5,.5){\circle*{.1}}
   \put(0,-.5){\circle*{.1}}
   \put(0,.5){\circle*{.1}}
   \put(.5,0){\circle*{.1}}
   \put(-.5,0){\circle*{.1}}
   \put(-.25,-.5){\vector(-1,0){0}}
   \put(.5,-.25){\vector(0,-1){0}}
   \put(.5,.25){\vector(0,-1){0}}
   \put(-.25,.5){\vector(1,0){0}}
   \put(.25,.5){\vector(1,0){0}}
   \put(-.5,-.25){\vector(0,1){0}}
   \put(-.5,.25){\vector(0,1){0}}
   \put(-.25,-.55){\makebox(0,0)[tr]{\footnotesize $\mu$}}
   \put(-.55,0){\makebox(0,0)[r]{\footnotesize $\nu$}}
   \put(0.1,-.55){\makebox(0,0)[t]{\footnotesize $x$}}
   \end{picture}}}

\newcommand{\twothree}{\setlength{\unitlength}{1cm}\raisebox{-.5cm}{
   \begin{picture}(1.2,1.2)(-.6,-.6)
   \put(-.5,-.5){\line(1,0){1}}
   \put(.5,-.5){\line(0,1){1}}
   \put(.5,.5){\line(-1,0){.5}}
   \put(-.5,.5){\line(0,-1){1}}
   \put(-.5,-.5){\circle*{.1}}
   \put(-.5,.5){\circle*{.1}}
   \put(.5,-.5){\circle*{.1}}
   \put(.5,.5){\circle*{.1}}
   \put(0,-.5){\circle*{.1}}
   \put(0,.5){\circle*{.1}}
   \put(.5,0){\circle*{.1}}
   \put(-.5,0){\circle*{.1}}
   \put(-.25,-.5){\vector(1,0){0}}
   \put(.25,-.5){\vector(1,0){0}}
   \put(.5,-.25){\vector(0,1){0}}
   \put(.5,.25){\vector(0,1){0}}
   \put(.25,.5){\vector(-1,0){0}}
   \put(-.5,-.25){\vector(0,-1){0}}
   \put(-.5,.25){\vector(0,-1){0}}
   \put(-.5,.7){\makebox(0,0)[tr]{\footnotesize $x$}}
   \put(-.55,0){\makebox(0,0)[r]{\footnotesize $\nu$}}
   \put(0.3,.7){\makebox(0,0)[t]{\footnotesize $\mu$}}
   \end{picture}}}

\newcommand{\twofour}{\setlength{\unitlength}{1cm}\raisebox{-.5cm}{
   \begin{picture}(1.2,1.2)(-.6,-.6)
   \put(-.5,-.5){\line(1,0){1}}
   \put(.5,-.5){\line(0,1){1}}
   \put(.0,.5){\line(-1,0){.5}}
   \put(-.5,.5){\line(0,-1){1}}
   \put(-.5,-.5){\circle*{.1}}
   \put(-.5,.5){\circle*{.1}}
   \put(.5,-.5){\circle*{.1}}
   \put(.5,.5){\circle*{.1}}
   \put(0,-.5){\circle*{.1}}
   \put(0,.5){\circle*{.1}}
   \put(.5,0){\circle*{.1}}
   \put(-.5,0){\circle*{.1}}
   \put(-.25,-.5){\vector(1,0){0}}
   \put(.25,-.5){\vector(1,0){0}}
   \put(.5,-.25){\vector(0,1){0}}
   \put(.5,.25){\vector(0,1){0}}
   \put(-.25,.5){\vector(-1,0){0}}
   \put(-.5,-.25){\vector(0,-1){0}}
   \put(-.5,.25){\vector(0,-1){0}}
   \put(0.25,.7){\makebox(0,0)[tr]{\footnotesize $x$}}
   \put(-.55,0){\makebox(0,0)[r]{\footnotesize $\nu$}}
   \put(-.25,.7){\makebox(0,0)[t]{\footnotesize $\mu$}}
   \end{picture}}}
\begin{document}
\begin{frontmatter}
\title{Anti-Ferromagnetic Condensate in Yang-Mills Theory}
\author[WUPP]{Jochen Fingberg\thanksref{DFG}} and
\author[STRA]{Janos Polonyi}
\address[WUPP]{Department of Physics,
               P.O. Box 10~01~27,
               University of Wuppertal,\\
               42097 Wuppertal,
               Germany}
\address[STRA]{Laboratory of Theoretical Physics,
               Louis Pasteur University,\\
               67084 Strasbourg,
               Cedex, France\\
               and\\
               Department of Atomic Physics,
               L. E\"otv\"os University,
               5-7 Puskin u.,\\
               Budapest 1088, Hungary}
\thanks[DFG]{J.F. is grateful for a fellowship from the
             Deutsche Forschungsgemeinschaft.}

\begin{abstract}
SU(2) gauge theory with competing interactions is shown to possess a
rich phase structure with anti-ferromagnetic vacua.
It is argued that the phase boundaries persist in the weak coupling
limit suggesting the existence of different renormalized continuum
theories for QCD.
\end{abstract}
\end{frontmatter}

\section{Introduction}
Non-renormalizable models are traditionally ignored in high energy
physics due to their lack of predictive power. In fact, one needs
infinitely many counter terms and renormalization conditions to render
these theories well defined.
In the framework of the renormalization group \cite{wilsonrg}
one usually identifies the ultraviolet fixed point, where the
correlation length diverges with the renormalized theory \cite{hasenrg}.
In the vicinity of the fixed point the renormalization group equations
can be linearized. The resulting scaling law allows to define
relevant, marginal or irrelevant operators. These are the
eigen-operators of the linearized renormalization group equation with
increasing, constant or decreasing coefficients as the cut-off is
lowered.
It is of central importance that the class of relevant and marginal
operators is equivalent to the class of renormalizable operators.
This can be seen from a comparison of two renormalized
trajectories, one with relevant or marginal couplings, the
other one with additional contributions from irrelevant operators. The
trajectories approach each other at the infrared end of the region of
linearizability. The difference between the two theories reduces to
an overall scale factor only as the cut-off is moved towards the
infrared. This shows that the presence of irrelevant operators
prevents the trajectory
to approach the fixed point as the cut-off is removed. In fact, if the
irrelevant coupling constant decreases in the infrared direction,
then it increases as the cut-off is enlarged.

Universality in the sense of independence on microscopical details
of the system implies that effects of irrelevant coupling
constants die out at physical scales. 
However, indications of continuum physics beyond the class of
traditionally renormalizable theories do exist.
One example to enlarge the class of theories, which characterize the
physics of a given particle and symmetry pattern, is based on the
presence of multiple fixed points \cite{senben}. 
Another, more conventional possibility exists, when strong anomalous
dimensions arise from non-perturbative effects. These effects have been
shown to change the sign of critical exponents from negative to positive
values \cite{miransky,bardeen}. For $RP^{N-1}$ models,
the role of non-perturbative topological
defect structures, which may survive the weak coupling limit and
influence the universality class of the theory, is under vigorous
discussion \cite{edwards,ballesteros,hasenbusch,niedermayer}.

In this paper we present indications that
the non-per\-tur\-ba\-tive generation of additional relevant operators
leading to competing interactions
might also be present in non-Abelian gauge theories.
Competing interactions are also relevant for
Yukawa models, where the existence of nontrivial fixed points
and the crossover between universality
classes has been investigated \cite{yukawa}.

We begin with a brief review of the basic mechanism,
the effect of higher order derivative terms, which are able to generate
competing interactions on the semi-classical vacuum.
In section 3 we discuss the possible phases of
the gauge models with higher derivatives in
the semi-classical picture. The Quantum Field Theoretical aspects
of the anti-ferromagnetic phase are briefly commented in Section 4.
Our lattice gauge model
is introduced in section 5. In the following chapter we present
numerical results and discuss the phase structure.
The vital issue of the continuum limit is addressed
in section 7. Finally we give a summary and conclude with
some speculative remarks.

\section{Non-Perturbative Effects at the Cut-off}
The renormalization group functions, which determine the effective
coupling constants are usually computed perturbatively.
There are some non-a\-symp\-to\-ti\-cal\-ly free models,
where strong interactions at the cut-off modify the
evolution of the coupling constants in a manner, which influences the 
long range structure of the vacuum. The formation of small positronium
bound states in strong coupling massless QED \cite{miransky} is an
example, where a strong anomalous dimension generates new relevant
operators \cite{bardeen}. Furthermore, the two dimensional Sine-Gordon
model has a phase transition at strong coupling, where the normal
ordering is not sufficient to remove the ultraviolet
singularities \cite{sinegr}.
Another description of this phase transition can be derived from the
equivalence of the Sine-Gordon and the X-Y model. It turns out that the
strong coupling phase of the Sine-Gordon theory is dual to the ionized
vortex phase of the X-Y model \cite{sineg}. In this phase the vortex
fugacity becomes a new relevant coupling constant.

We want to address the question of condensate formation at the cut-off
scale in non-Abelian gauge models. For simplicity we consider an
SU(2) gauge model with additional higher derivative terms in the action.
It is defined by the Lagrangian
\be
{\cL}=
-{1\over4g^2}~F^a_{\mu\nu}\biggl(\delta^{ab}+{c_2\over\Lambda^2}\cD^{2ab}
+{c_4\over\Lambda^4}\cD^{4ab}\biggr)F^{b\mu\nu}~~,
\label{eq:conac}
\ee
where $\cD^{ab}_\mu$ is the covariant derivative.
There are two different ways of looking at this modified theory.
One possibility is to start with a theory of elementary particles.
In this case the cut-off has to be removed and the
 $\Lambda$-parameter must be identical to the cut-off
in order to eliminate overall UV divergences.
The new terms in the action can be regarded as a variant
of the Pauli-Villars regulator.
The theory is renormalizable in the framework of a perturbative
expansion in $g^2$ and $c_\alpha$ around the
configuration with $A_\mu(x)=0$ \cite{rieszqcd,zinn}, provided
the one-loop level is regulated \cite{slavnov}.
Another possibility is to consider a unified model,
where a heavy particle
is coupled to a Yang-Mills field. The elimination of the heavy particle
generates new terms in the Lagrangian of eq.\ref{eq:conac}.
Consider a sequence of theories in particle physics, ${\cal T}_k$ with
characteristic energy scales $\Lambda_k$, $\Lambda_{k+1}>\Lambda_k$
($k=0,1,\cdots,N$), where ${\cal T}_N$ 
is the Theory of Everything (TOE) and ${\cal T}_k$
is an effective theory of ${\cal T}_{k+1}$ with cut-off
${\cal O}(\Lambda_{k+1})$. 
The particle exchange at the 
characteristic scale of ${\cal T}_{k+1}$ generates relevant
and irrelevant operators for ${\cal T}_k$ according to the
decoupling theorem \cite{appelq}. 
Though irrelevant couplings are suppressed as 
${\cal O}(\Lambda_k^2/\Lambda_{k+1}^2)$, they are important
because they represent
physics beyond the scale $\Lambda_k$. In their absence high
energy scaling would be described by ${\cal T}_k$ and the renormalized
trajectory would end at the UV fixed point of ${\cal T}_k$. These
irrelevant coupling constants are the physical
regulators of ${\cal T}_k$.
In this case no attempt is made to remove the cut-off of this
effective theory.

An observable $A$ at momentum scale $\mu$ has a perturbative expansion:
\be
A=\mu^{[A]}\sum_{klm}I_{klm}~g^k
\biggl(c_2{\mu^2\over\Lambda^2}\biggr)^l~
\biggl(c_4{\mu_4\over\Lambda^4}\biggr)^m~~,
\label{eq:pertir}
\ee
where $I_{klm}$~ is a dimensionless loop integral. In the absence of
infrared singularities observables are independent of $c_2$ and $c_4$
far away from the cut-off. This demonstrates the irrelevance of 
coupling constants with negative mass dimension.
The important question now is, whether this argument, which is based on
an expansion around the trivial vacuum, $A_\mu=0$, remains valid in the
presence of an inhomogeneous condensate. 

To answer this question, first consider the action corresponding to an
instanton with scale parameter $\rho$ \cite{enzo}
\be
S_{inst}(\rho)={8\pi^2\over g^2}~
    \biggl(1-{\tilde c_2\over(\rho\Lambda)^2}
+{\tilde c_4\over(\rho\Lambda)^4}\biggr)~~,
\label{eq:insta}
\ee
where $\tilde c_\alpha=O(c_\alpha)>0$. This action agrees with the
renormalized
value, $8\pi^2 / g^2$, for large instantons. As the instanton size 
approaches the length scale $\Lambda^{-1}$ we find deviations from
scale invariance and a minimum is reached
in the vicinity of the cut-off,
\be
S_0=S(\rho_0)
   ={8\pi^2\over g^2}\biggr(1-{\tilde c_2^2\over4\tilde c_4}\biggr)~~,
\label{eq:mact}
\ee
where
\be
\rho_0=\Lambda^{-1}\sqrt{{2\tilde c_4\over\tilde c_2}}~~~~.
\ee
Since the mini-instanton action is not a simple polynomial
of the coupling 
constants it generates a combination of $c_2$ and $c_4$,
which is independent
of the subtraction scale $\mu$ and the result depends
on $c_\alpha$ in contrast to eq.\ref{eq:pertir}. The absence of the
usual power suppression in the infrared region opens the possibility
of an eventual modification of the universality class.

In order to find a regime of the coupling constants, where the
mini-in\-stan\-tons dominate we increase $c_2$.
At some critical value $c_0$ of $c_2$
the mini-instanton gas will dominate the path integral. Beyond this 
point $c_2 > c_0$ we find negative values of the action
given by eq.\ref{eq:mact}. This is a radically new territory.
The negative instanton action imposes frustration on the vacuum.
It becomes energetically favorable to fill the vacuum with 
instantons. The resulting semi-classical ground state will be a
crystal of instantons. For small $g^2$ the instanton lattice is densely
packed and lattice vibrations are strongly suppressed.
The translation and rotation invariance of this semi-classical 
vacuum will be broken just as in the case of an ionic lattice in
solid state physics. As long as CP symmetry is preserved the total
winding number of the vacuum vanishes. Then the crystal consists of
alternating instantons and anti-instantons similar to an 
anti-ferromagnetic (AF) N\'eel state.

The matching between the topology of internal
and external spaces is
not essential in establishing the AF order, in contrast with
discrete spin models. In fact, the eigenvalue of small
fluctuations around
 $A_\mu=0$ with momentum $p$ in Feynman gauge is given by
\be
\lambda(p^2)=p^2\biggl(1-c_2{p^2\over\Lambda^2}+
                         c_4{p^4\over\Lambda^4}\biggr)~~.
\ee
The instability, $\lambda(p^2)<0$, occurs for momenta
\be
{1\over2}-\sqrt{{c_2^2-4c_4\over4c_4^2}}<{p^2\over\Lambda^2}<
{1\over2}+\sqrt{{c_2^2-4c_4\over4c_4^2}}~~~~.
\ee
When $c_2>2\sqrt{c_4}$ this leads to a condensation of particles in this
momentum range until their repulsive interactions stabilize the vacuum.
The result will be a condensate with staggered order parameter
in the middle of the Brillouin-zone, $p\approx\Lambda/2$.

\section{Phase Structure}
One expects three qualitatively different
phases shown in fig.\ref{fig:f1} for a continuum 
theory with the Lagrangian of eq.\ref{eq:conac}: 
\begin{figure}[htb]
\centering\epsfig{file=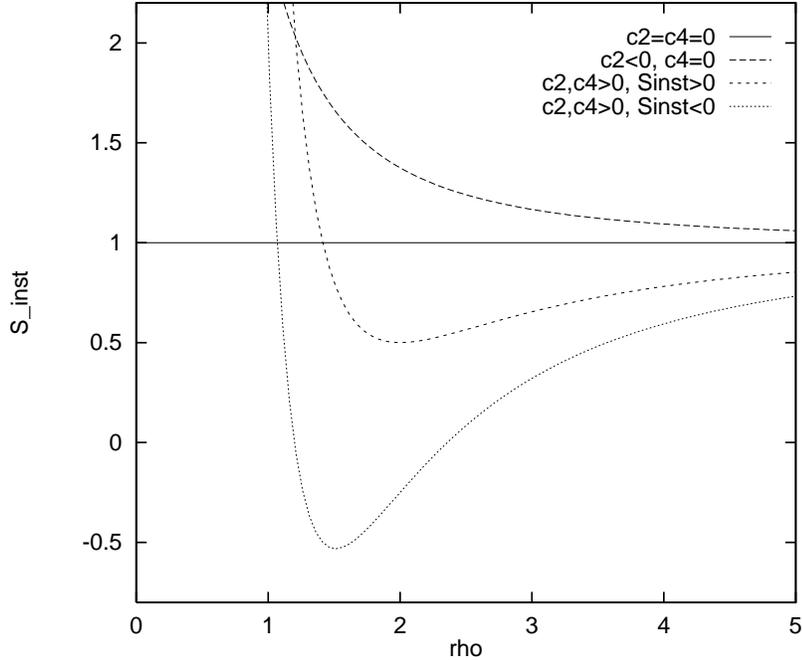,width=0.8\linewidth}
\caption{
Size dependence of the instanton action in the three different phases
of the gauge theory. Full and long dashed lines correspond to
$c_\alpha=0$ and $c_2<0$, respectively, and belong to phase 1.
The short dashed line stands for $c_2>c_0$, $c_4>0$,
which corresponds to
phase 2. The AF phase 3 is shown by a dotted line.}
\label{fig:f1}
\end{figure}

\begin{enumerate}
\item {\it Classically scale invariant phase:}
$c_\alpha=0$. The theory should be in the same universality
class as the one with $c_2<0$ and $c_4>0$. For this
choice of the new coupling constants, $c_\alpha$, the saddle
points, i.e. $A_\mu=0$ and instantons with cut-off independent
size parameter are not influenced by $c_\alpha$ and remain in the
infrared. As a consequence the usual
argument for the irrelevance as explained in
context of eq.\ref{eq:pertir}
applies. This is the usual phase of the Yang-Mills system known from
dimensional regularization. 

\item {\it Mini-instanton phase:} $c_4>0$, $c_0>{c^2_2\over c_4}>0$
and $S_0$ is small and positive. 
The path integral is dominated by instantons with size
close to the cut-off. 
The saddle points in the ultraviolet regime
generate different beta-functions and the physics changes 
discontinuously between phases one and two. 

\item {\it Anti-ferromagnetic phase:} ${c^2_2\over c_4} > c_0$ so 
that $S_0<0$. The
semi-classical vacuum is an instanton lattice with alternating
topological charge. One finds this reminiscent of the N\'eel state of
solid state physics at $p\approx\Lambda/2$. Further splitting into 
sub-phases without anti-ferromagnetic long range order is possible.
Renormalizability is a highly non-trivial issue.
\end{enumerate}

The usual attitude towards mini-instantons, alias topological defects
\cite{topd}, is to suppress them either by improving the action
\cite{impra}, or the definition of the winding number
\cite{impro}, in order to stay in phase one.
The attempt to eliminate these modes originates
from the naive semi-classical picture, where the configurations
that dominate the path integral are solutions of the
renormalized, cut-off independent equations of motion.
But this is not necessarily so \cite{schulman}.
The d-dimensional lattice regulated path integral is saturated by 
configurations, whose variation from site-to-site is
\be
\Delta\phi(x)=O(a^{1-{d\over2}})~~~~.
\label{eq:nscal}
\ee
This result is the real space equivalent to the momentum space
power counting rule. For a free massless theory it can be derived
from the path integral
\be
\prod_x\int d\phi(x)~~
\mbox{e}^{-{1\over2}a^{d-2}\sum_{x,\mu}(\phi(x+\mu)-\phi(x))^2}~~~~.
\ee
The Gaussian integration yields eq.\ref{eq:nscal}, which
leads to continuous but nowhere differentiable
trajectories in quantum mechanics for $d=1$ \cite{rqm}, 
finite, cut-off independent discontinuities and non-trivial
phase structure 
for $d=2$ \cite{sineg}, and Dirac delta singularities
in higher dimensions \cite{paris}. It is a complicated dynamical
issue, whether smoothness prevails for certain observables in quantum
field theory.

Recent improvements \cite{perfect}, go so 
far as to try to cancel all power corrections to the scale
invariant action.
This certainly brings the improved lattice theory close to the
renormalized trajectory of perturbative QCD known from dimensional
regularization. Our strategy is just the opposite. We
exploit the richness of gauge theories by varying the
dynamics close to the cut-off and see, whether non-perturbative
effects generate new quantum field theories.

It has been known that competing interactions
create rich phase structure. An AF phase is found,
where frustrations stabilized by the balance of repulsive and
attractive forces form a densely packed lattice.
Systems showing frustration
have first been investigated in the framework of solid state physics
\cite{selke},
where it was found that the AF phase usually consists of
several layered sub-phases. Recent results obtained for the four
dimensional Ising model \cite{zk,parisi}, the
three dimensional $Z_3$ \cite{zh} Potts model
with ferromagnetic (FM) nearest neighbour (NN) and AF next to nearest
neighbour (NNN) couplings can be considered as generic.
Starting in the FM phase and increasing the strength of the AF coupling
we enter a phase with AF order in one or two directions.
Further increase of the AF coupling
induces more directions to show AF behaviour.
Completely AF order may or may not be reached 
for strong AF coupling depending on the matching
of space-time and the internal dimensions. The AF
phase has already been considered in ref.~\cite{gallavotti} where 
perturbative arguments for the existence of a non-trivial UV limit of the
lattice $\phi^4$ theory were found.

\section{Anti-Ferromagnetic Vacuum}
The use of AF condensates in Quantum Field Theories
deserves special attention
due to additional difficulties with
reflection positivity and the construction of the continuum limit.
When eq.\ref{eq:conac} is considered as an
effective theory obtained from the perturbative elimination of a particle
with mass $M={\cal O}(\Lambda)$, the cut-off, $\Lambda$, is kept finite.
This model has two phases, the usual one with $S_0>0$ and a frustrated phase
where $S_0<0$.
The length scale of the frustrated vacuum, i.e.{} the lattice spacing of the 
instanton-anti instanton crystal, ${\cal O}(\Lambda^{-1})$, is kept finite.
It is difficult to describe the physics beyond the energy $\Lambda$,
the heavy particle threshold, by an effective theory without
the heavy particle as a dynamical degree of freedom.
In this energy regime the propagating heavy particle and
its open channels make the truncation
of the effective action in eq.\ref{eq:conac}
to ${\cal O}(\Lambda^{-4})$
a very poor approximation.
To point out a systematic approach to this problem we remark that
the time evolution is unitary in a positive definite Hilbert space
for a well defined Quantum Field Theory. In the Euclidean domain
this translates to reflection positivity \cite{posit}.
A sufficient condition for reflection positivity of the single plaquette action
$S=\sum_j c_j\sum_{x,\mu\nu}\mbox{Tr}_j(1-U_{\mu\nu}(x))$ where
$\mbox{Tr}_j$ denotes the trace in a representation labeled by $j$
is that the coefficients of the traces are positive for each representation.
The lattice regulated version of
eq.\ref{eq:conac} contains negative coefficients 
in front of the traces in the Lagrangian and the issue of the reflection 
positivity is unclear.

We believe that the problem with unitarity has a deeper origin.
In a renormalization procedure a blocking step unavoidably generates
coupling constants with negative sign. These couplings represent
interactions between remote sites on the lattice. Does this mean that the
concept of unitarity or reflection positivity is not renormalization group
invariant ? This might be so because even the transfer matrix formalism is
lost after a blocking step due to the presence of 
higher order derivatives in the action. When degrees of 
freedom are eliminated in a quantum system pure states become
mixed states and one should use the density matrix formalism. Higher order 
derivative terms responsible for
the loss of the transfer matrix formalism represent correlations
which are usually taken into account by the density matrix formalism.
As long as they are irrelevant this mixing of different quantum states is 
negligible. In fact, in the ultraviolet scaling regime of the trivial
or ferromagnetic vacuum where higher order derivatives are certainly
irrelevant the effects of heavy modes eliminated by the
blocking step are local and a violation of unitarity is observable
only at short distances.

This lets us speculate that a generalization of unitarity or reflection
positivity to AF systems is possible by
relaxing these conditions in the ultraviolet region and demanding
their strict validity only at physical length scales, i.e.{} in
the continuum limit. 

It has been long known that higher order derivatives can generate new degrees
of freedom with negative norm which violate unitarity \cite{pais}.
The violation of unitarity can be understood looking at the frequency
integration in perturbative loop integrals. Higher powers of the
momentum in the denominator lead to poles at complex frequencies
giving non-unitary contributions. The usual strategy
to recover unitarity is to fine tune the parameters of the theory so that
the poles appear in complex conjugate pairs and the imaginary part of their
contribution cancel \cite{leewick,kuti}. No violation of unitarity is
detected at finite energy in the perturbative solution when the scale
of the coupling constants of higher order derivatives, $\Lambda$,
approaches infinity. Such a solution seems to support our
suggestion above, namely to relax the strict conditions imposed at all scales.
However, the vacuum considered in this paper is non-homogeneous and the argument
of ref.~\cite{leewick,kuti} is not sufficient to exclude the violation 
of unitarity.

We think that the problem with unitarity and reflection positivity
can be solved for the AF vacuum. The AF
phase has actually been observed in Solid State Physics for quantum systems.
How does the anti-ferromagnetic Heisenberg model retain reflection
positivity despite the negative sign of its coupling constant? An illuminating
example is worked out in ref.~\cite{chung} where the quantum
mechanics of a $Z_4$ variable is considered. It is found that the transfer
matrix, $T$, may have real but negative eigenvalues when the coupling is
"anti-ferromagnetic" in time. It is suggested to impose reflection
positivity for the square of the transfer matrix, $T^2$, only. Note
that $T^2$ describes the jump in time by the period length
of the "anti-ferromagnetic" vacuum in time. This agrees
with our intuition: The AF vacuum reflects the presence of
periodic structures in space or time. There should be
no problem with unitarity if
\begin{itemize}
\item
the vacuum is static, "ferromagnetic", or
\item
only those powers of the transfer matrix
are taken into account whose time step is an integer multiple 
of the period length.
\end{itemize}
As a slight generalization of this conjecture we
expect that the renormalized trajectory arising from a blocking 
procedure which preserves the AF order, 
\cite{afisrg}, maintains reflection positivity.

The construction of the continuum limit, $\Lambda\to\infty$, of
eq.\ref{eq:conac} raises the question of renormalizability.
Although not required for realistic, i.e.{} effective theories,
it simplifies ultraviolet scaling laws and 
eliminates the unphysical complications at high energies
in the anti-ferromagnetic vacuum.
However, in Quantum Field Theories with AF vacuum
renormalizability has
another important role. The inhomogeneous vacuum violates conservation 
laws related to external symmetries. The space-time symmetries
of the continuum Quantum Field Theory are recovered as the length scales
of both the anti-ferromagnetic condensate and the cutoff vanish.

The renormalized theory is defined at the critical point,
where at least one of the correlation lengths of different particle
sectors diverges. The selection of one sector with diverging 
correlation length and keeping the corresponding particle mass
constant fixes the value of the cutoff. Other sectors, where
the correlation length diverges slower or faster when the critical
point is approached correspond to infinitely heavy
or massless excitations, respectively. The physical content of
the theory in this limit can depend on the direction in which the
critical point is approached.

The continuum limit of an AF vacuum is a rather involved problem.
Analytical methods based on an expansion around
homogeneous or slowly varying field configurations are inadequate for its
study. In fact, in the presence of an AF condensate
the plaquettes differ from a constant configuration
by a finite amount, so that $<A_\mu>=\Lambda O(g^{-1})$.
Such an extremely strong field can modify the 
UV scaling laws verified rigorously in ref.~\cite{magriv}.

The particle content of a theory can be studied by the help of
Bloch-waves. The excitations above a
vacuum which has $N_0$ degrees of freedom in an elementary cell
give rise to $N_0$ dispersion relations, the bands in the language of Solid
State Physics. There might be several particles corresponding to
a single quantum field. 
Some of the bands may correspond to excitations whose mass
diverges with the cut-off. These heavy particles decouple from the physics
during the renormalization procedure. The others are renormalized and
describe particle-like excitations with finite mass.

The renormalized AF vacuum possesses a condensate whose 
characteristic length approaches
zero and becomes unresolvable for observables
which are defined at finite scales. 
As a consequence, both the AF and the ferromagnetic vacuum appear homogeneous
after renormalization. What distinguishes these phases are the excitations
above the vacuum.
The ultraviolet fixed point
was studied in ref.~\cite{vega} when only one particle mass was kept finite
during the renormalization. The situation is similar to lattice fermions
which form an anti-ferromagnetic vacuum \cite{semenoff}.
If some of the particle modes acquire a mass proportional to the
cut-off a generalization of the decoupling theorem, \cite{appelq,vega}, 
to anti-ferromagnetic vacua can be used to describe their effects on 
the particles remaining after renormalization. The
renormalization can be studied in a perturbative
expansion even when the mass of several particles is kept finite
\cite{herve}.

One expects new universality classes in the AF phase
because the interactions between different particle modes of the 
same field operator are described by vertices which are delocalized
within the elementary cell of the vacuum. A study of the two dimensional 
Ising model revealed new critical points and relevant operators in the
AF phase compared to the disordered
and ferromagnetic phases \cite{afisrg}. The negative eigenvalue of the transfer 
matrix mentioned above introduces a staggered mode at arbitrary 
distance. This gives an example how the ultraviolet structure of the vacuum may
modify long range correlations.

\section{Lattice Regulated Model}
The lattice regulated version of the irrelevant operator
$F^a_{\mu\nu}\cD^{2n\/ab}F^{b\mu\nu}$ corresponds to a linear
superposition
of Wilson loops up to size $2n\times2n$. Although it is straightforward
to find a lattice version of the action defined in
eq.\ref{eq:conac}, we choose an even simpler, more local form
of the lattice action.
In fact, we do not have to rely on the tree level stability
of instantons in a fully non-perturbative numerical computation.
It is sufficient to use a
modified action, which includes negative contributions from
short scale fluctuations \cite{gallavotti}. This can 
be achieved with an extended lattice action,
which contains planar $1\times1$, $1\times2$ and $2\times2$ loops,
\bea
S=&c_1& S_{1x1} + c_2 S_{1x2} + c_3 S_{2x2}\nonumber\\[\baselineskip]
 =&c_1&\sum_{x,\mu > \nu}\Tr\left(1-\plaq\right)\nonumber\\
 +&c_2&\sum_{x,\mu > \nu}\left[~\Tr\left(1-\twooneplaq\right)
 +                              \Tr\left(1-\onetwoplaq\right)\right] \nonumber\\
 +&c_3&\sum_{x,\mu > \nu}\Tr\left(1-\twoplaq\right) \quad,
\label{eq:Sext}
\eea
where our notation follows the definitions of ref.~\cite{LW}.
For completeness, we present the dependence of the action on a
specific link.
The action per link $U_\mu(x)$, which is needed in the
heat-bath algorithm, is given by
\be
S(U_\mu(x))=2~\RE\Tr[1-U_\mu(x)\tilde{U}_\mu(x)]\quad.\label{eq:Su}
\ee
It contains the sum of the staples associated with the Wilson loops
appearing in the extended action, eq.\ref{eq:Sext},
\bea
\tilde{U}_\mu(x)=\sum_{{\nu}\atop{\nu \ne \mu}}
 &c_1&\left(~\stapup+\stapdw ~\right)\nonumber\\[1.2\baselineskip]
+&c_2&\left(\wplaqone+\wplaqtwo+
 \wplaqthree+\right.\nonumber\\[1.2\baselineskip]
 &{ }&\left.~~\wplaqfour+\wplaqfive+
              \wplaqsix\right)\nonumber\\[1.2\baselineskip]
+&c_3&\left(\twoone+\twotwo+\twothree+\twofour\right)\quad.
\label{eq:stap}
\eea

\noindent
The classical continuum limit yields the operator content
\cite{LW,Morningstar}
\bea
 S_{1x1} &=& -\frac{g^2}{2N} \sum_{\mu > \nu}~[  \frac{a^4}{2}~\Tr F_{\mu\nu}^2
               - \frac{ 1}{12} a^6 ~\Tr(\cD_\mu F_{\mu\nu})^2
               + {\cal O}(a^8) ~] \\
 S_{1x2} &=& -\frac{g^2}{2N} \sum_{\mu > \nu}~[ 8\frac{a^4}{2}~\Tr F_{\mu\nu}^2
               - \frac{ 5}{ 3} a^6 ~\Tr(\cD_\mu F_{\mu\nu})^2
               + {\cal O}(a^8) ~] \\
 S_{2x2} &=& -\frac{g^2}{2N} \sum_{\mu > \nu}~[16\frac{a^4}{2}~\Tr F_{\mu\nu}^2
               - \frac{16}{ 3} a^6 ~\Tr(\cD_\mu F_{\mu\nu})^2
               + {\cal O}(a^8) ~]
\label{eq:Sclassical}
\eea
giving the sum
\bea
S=&-&\frac{g^2a^4}{~4N}~(c_1+~8c_2+16c_3)~\sum_{x,\mu > \nu}
    ~\Tr F_{\mu\nu}^2\nonumber\\
  &+&\frac{g^2a^6}{24N}~(c_1+20c_2+64c_3)~\sum_{x,\mu > \nu}
    ~\Tr(\cD_\mu F_{\mu\nu})^2
   +{\cal O}(a^8)~~.
\label{eq:extac}
\eea
Note that the action in eq.ref{eq:extac} is bounded from below 
by the terms ${\cal O}(a^8)$ for any sign of the coupling constants.
The condition of having a given value of $g^2$ is
\be
c_1 + 8 c_2 + 16 c_3 = 1
\label{eq:beta}
\ee
Naively the region of stability of negative plaquettes is given by
\be
c_1 + 3 c_2 + 2 c_3 < 0
\label{eq:stab}
\ee
as can be seen in eq.\ref{eq:stap} from the number of
staples attached to a given link.
At tree level the ${\cal O}(a^6)$ terms can be removed from
the action eq.\ref{eq:extac} \cite{Symanzik,LW1} if
\be
c_1 + 20 c_2 + 64 c_3 = 0 ~~~~.
\ee
Commonly used examples \cite{Morningstar,thermo,improved}
of Symanzik improved actions are found for the combinations
\bea
S_{(1,2)}~&=&~\frac{5}{3} S_{1x1} - \frac{1}{12} S_{1x2} ~~~,\\
S_{(2,2)}~&=&~\frac{4}{3} S_{1x1} - \frac{1}{48} S_{2x2} ~~~.
\eea

\noindent
To continue it is convenient to change to a new coordinate system
\bea
c\left(u,\rho,\phi\right)_\alpha ~&=&~
\hat u_\alpha+\rho ~(\cos\phi~
\hat v_\alpha+\sin\phi ~\hat w_\alpha)\\[1.2\baselineskip]
\hat u_\alpha&=&( 1,  8,16)/\sqrt{  321}\nonumber\\
\hat v_\alpha&=&( 8, -1, 0)/\sqrt{   65}\nonumber\\
\hat w_\alpha&=&(16,128,-9)/\sqrt{16721}\nonumber
\eea
in a coupling space with one classically relevant direction,
$\hat u_\alpha$, and two
classically irrelevant directions, $\hat v_\alpha$ and $\hat w_\alpha$.
However, one should bear in mind that the
division of the operator content 
of the action defined in eq.\ref{eq:extac}
\be
S~=~[\hat u_\alpha + \hat v_\alpha + \hat w_\alpha ]~\cdot~
    (S_{1\times 1},S_{1\times 2},S_{2\times 2})
\ee
into relevant and irrelevant contributions
is meaningful only in the usual FM phase
corresponding to phase one of section~3
due to contributions ${\cal O}(a^8)$,
which become important in the AF phase.

\section{Numerical Results}
AF ordering is expected to show up as fast oscillating behaviour
of certain correlation functions. A positive term in the action
with extension of $n$ lattice spacings and a negative coefficient
may generate AF short range order with period
length up to $n$ lattice spacings even in the disordered phase.
Thus, our simple action is expected to possess an AF
ground state with period length of two lattice spacings.

A simple indicator for AF ordering can be defined by gauge
invariant correlation function $\Gamma_{\mu\nu\alpha\beta}^{\rho}(x)$
of elementary plaquettes,
\be
\Gamma_{\mu\nu\alpha\beta}^{\rho}(x_n=n~\hat\rho)=<S_{\mu\nu}^{1\times 1}(0)~~
                                  S_{\alpha\beta}^{1\times 1}(n~\hat\rho)>~~.
\label{gamma}
\ee
The function $\Gamma_{\mu\nu\alpha\beta}^{\rho}(x)$ measures
correlations of plaquettes $S_{\mu\nu}^{1\times 1}$
orientated in the $\mu\nu$-plane along the $\rho$-axis.
As a 2-dimensional example think of a checkerboard of
positive and negative plaquettes. This configuration has 2 AF 
directions. In general, different layered sub-phases
are characterized by the number of AF directions.
However, one should bear in mind that this number may not identify
the sub-phase uniquely, since different AF rearrangements
are possible in a given set of planes.

\begin{figure}[htb]
\centering\epsfig{file=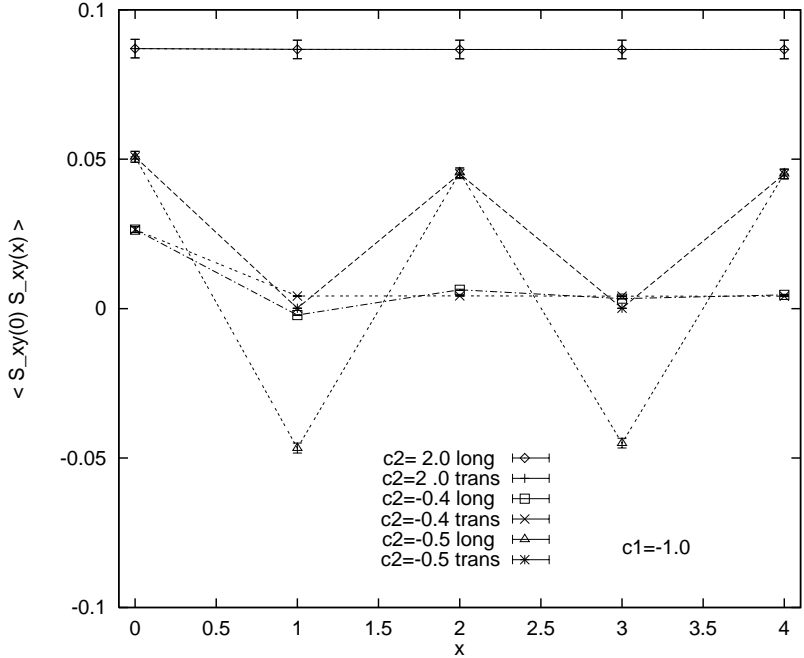,width=0.8\linewidth}
\caption{Longitudinal and transverse correlation functions
         of xy-plaquettes for $c_2=2.0$ (FM),$-0.4$ and $-0.5$ (AF),
         $c_3=1/16-c_1/16-c_2/2$ and $\beta=2.4$.}
\label{fig:f2}
\end{figure}

\begin{figure}[htb]
\centering\epsfig{file=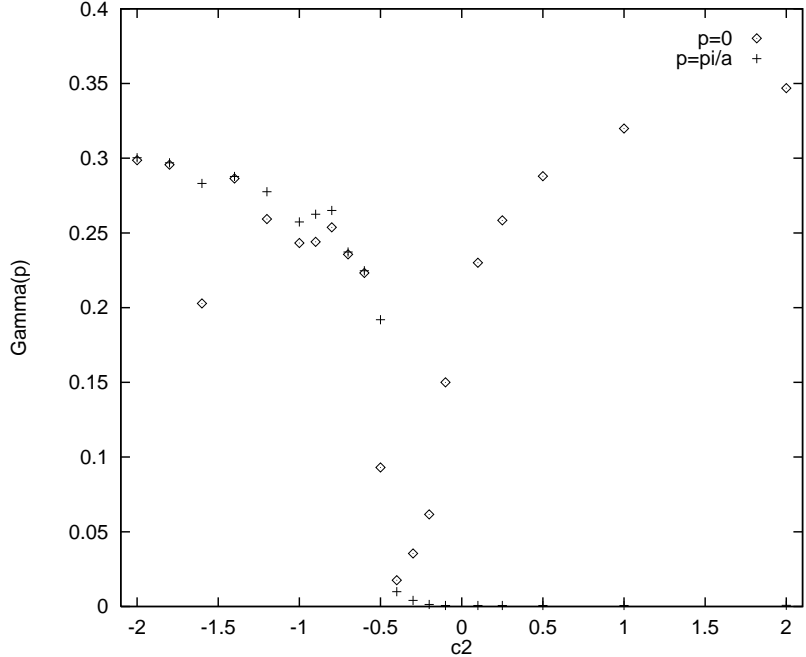,width=0.8\linewidth}
\caption{Normal ($p=0$) and staggered ($p=\pi/a$)
         order parameter as a function of $c_2$
         for $\beta=2.4$, $c_1=-1$, and $c_3=1/16-c_1/16-c_2/2$.}
\label{fig:f3}
\end{figure}

\begin{figure}[htb]
\centering\epsfig{file=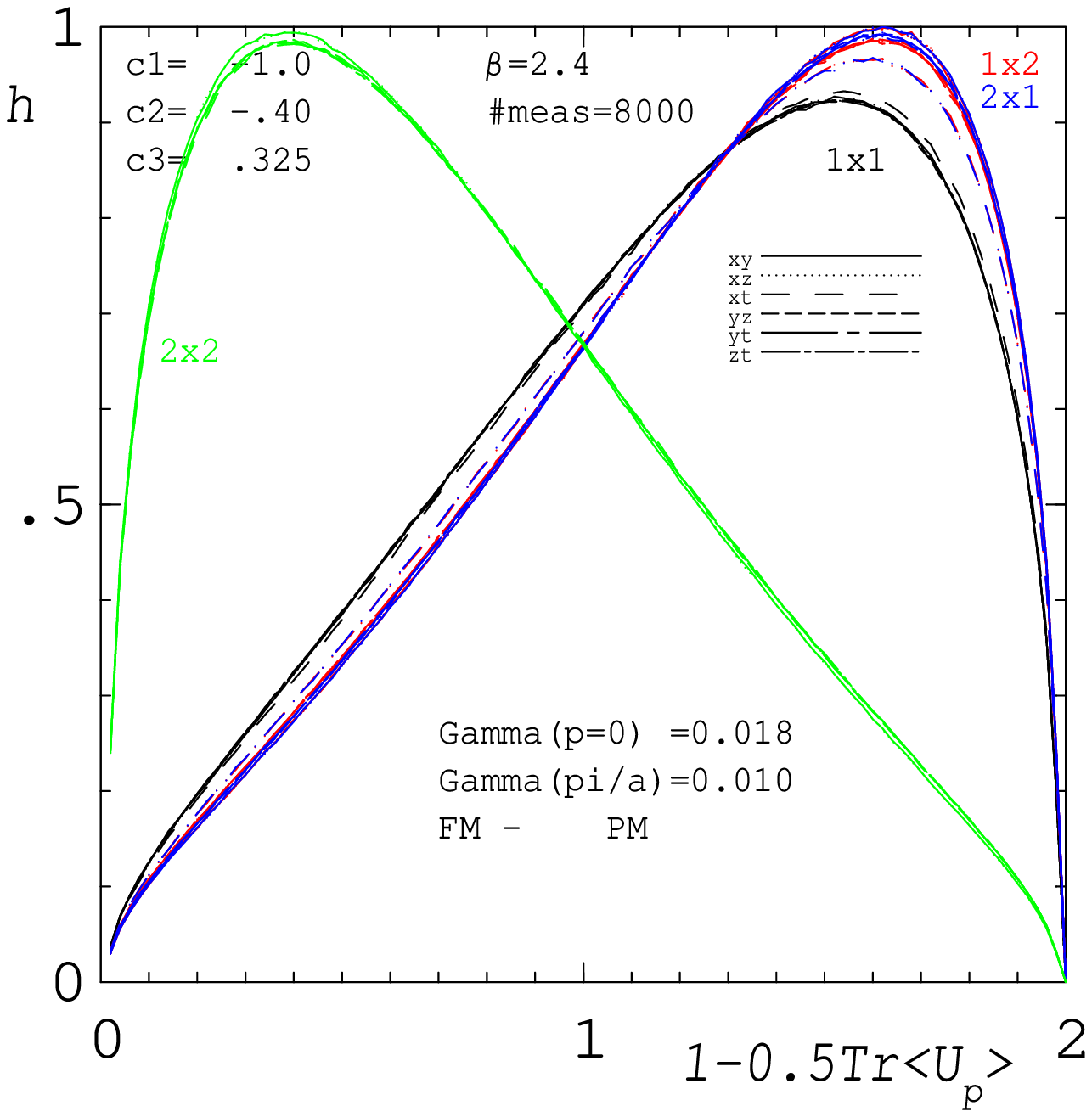,width=0.8%
\linewidth,bbllx=100,bblly=200,bburx=500,bbury=600}
\caption{Histogram of Wilson loops in the action, $c_2=-0.4$.}
\label{fig:f4}
\end{figure}

\begin{figure}[htb]
\centering{\epsfig{file=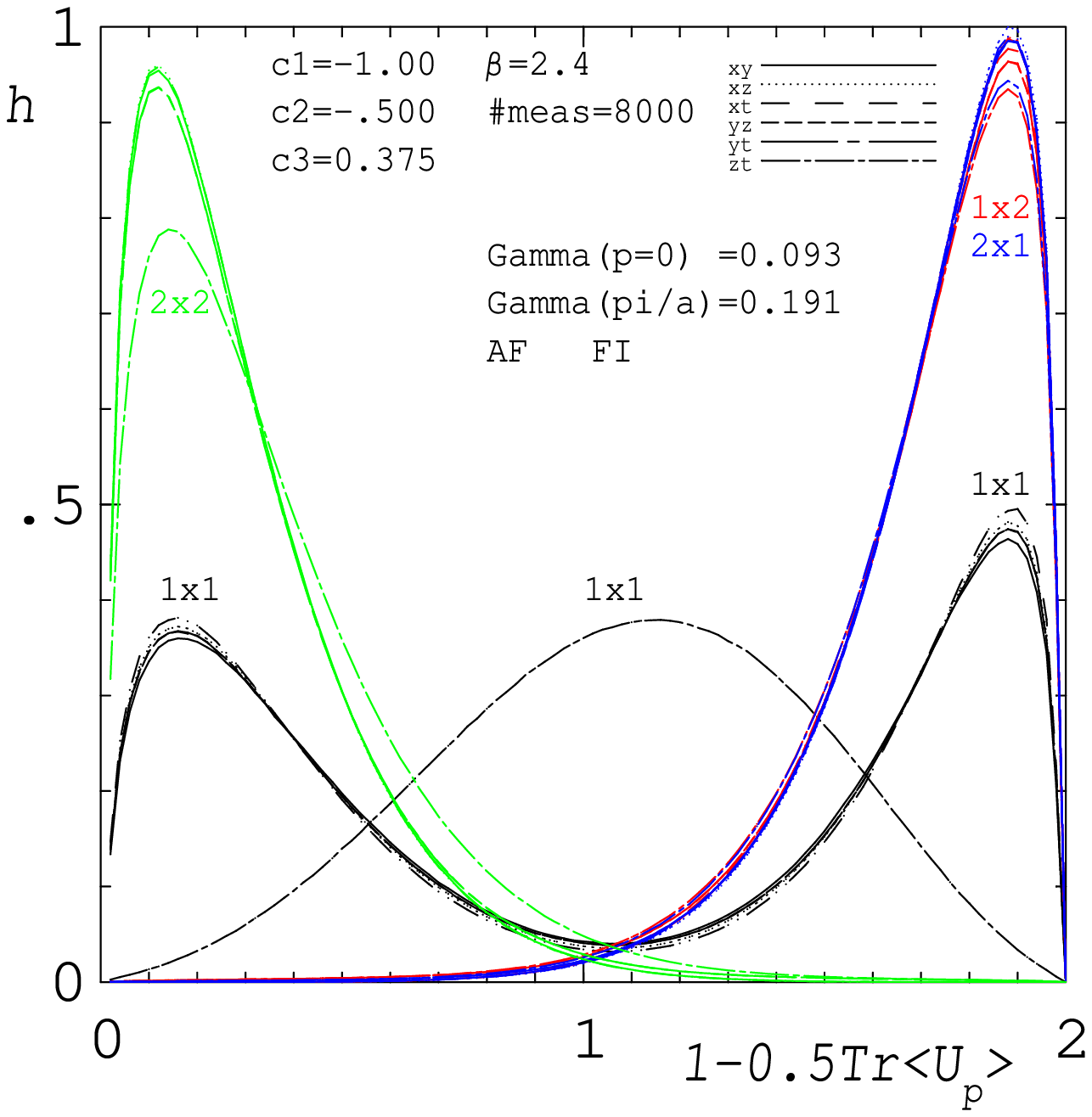,width=0.8%
\linewidth,bbllx=100,bblly=200,bburx=500,bbury=600}}
\caption{Histogram of Wilson loops in the action, $c_2=-0.5$.}
\label{fig:f5}
\end{figure}

$\Gamma_{\mu\nu\alpha\beta}^{\rho}(x)$ can be seen in fig.\ref{fig:f2},
where the longitudinal
($\hat\rho\cdot\hat\mu=1$ or $\hat\rho\cdot\hat\nu= 1$)
and transverse ($\hat\rho\cdot\hat\mu=\hat\rho\cdot\hat\nu=0$)
correlation functions $\Gamma_{\mu\nu\alpha\beta}(x)$ defined in eq.\ref{gamma}
are shown in the FM and AF phase.
The lattice size used in the computation was $8^4$ and the coupling
constant was chosen to be $\beta=2.4$. When a range of the coupling
constant $c_2$ is studied $c_3$ is always adjusted according to
eq.\ref{eq:beta} to keep $\beta$ fixed.
An AF phase is expected when one of the
$c_\alpha$ becomes sufficiently negative and the resulting frustrations
drive the system to develop oscillatory $\Gamma_{\mu\nu\alpha\beta}^{\rho}(x)$.
A convenient order parameter for the AF transition
is the Fourier transform of the plaquette-plaquette correlation 
function at the cut-off,
\bea
\Gamma_{\mu\nu\alpha\beta}^{\rho}(p_l) &=& \sum_{j=0}~
                   \Gamma_{\mu\nu\alpha\beta}^{\rho}(x_j)
                                          ~\exp{(\/i~p_l~x_j)}\\
                      p_l &=& 2\pi l~/~L\nonumber\\
                      x_j &=& jL~/~N\nonumber ~~,
\eea
reminiscent of the staggered magnetization for spin models.
Here $L=Na$~ is the length of the lattice with spacing $a$.
An AF condensate can be found as a maximum of the
Fourier transform of the correlation functions
at the upper edge of the Brillouin-zone, $p=\pi/a$.
One has to keep track of the transverse and
longitudinal correlation functions separately.
We found that longitudinal correlation functions displayed the
AF order earlier and with larger amplitude. 
Since it is impossible to predict, in which plane the AF
ordering will take place we consider the maximum of the correlation
functions
\be
\Gamma(p)=\max_{\mu\nu\alpha\beta,\rho}~\mid~\Gamma_{\mu\nu\alpha\beta}^{\rho}
    (p)~\mid~~.
\ee
The dependence of $\Gamma(p)$ on $c_2$ is shown in fig.\ref{fig:f3}.
An abrupt change both for $p=0$ and $p=\pi/a$
suggests the existence of a transition in the
region of $c_2\approx -0.4$. We see FM ordering above this
point and AF or ferrimagnetic (FI) ordering below. 
The number of AF space-time directions can be read
from the distributions of different Wilson loops in the action
shown in figs.\ref{fig:f4} and \ref{fig:f5}.
In fact, FM or AF order corresponds
to single or double peaked histograms for plaquettes with
orientation $\mu\nu$.
One expects that the AF semi-classical vacuum
can be constructed of link variables, which are elements of
the center of the gauge group, analogous to a N\'eel state.
This type of configuration satisfies the lattice equations of
motion and saturates the action in the AF phase.
However, in case of a mismatch between internal and external
topology the competing interactions can be balanced by forming
incommensurate structures \cite{selke} 
and the semi-classical vacuum will become
even more complicated.
The histogram in fig.\ref{fig:f4} shows that FM
Wilson loops are concentrated around $\pm 1$ up to quantum fluctuations.
This suggests a $Z_2$ valued semi-classical vacuum.
On the contrary, AF planes also have 
non-center link variables $U\approx e^{i\pi\hat n_j\sigma_j/2}$,
$\hat n_j\cdot\hat n_j=1$ demonstrated in fig.\ref{fig:f5}
and the semi-classical vacuum shows a non-Abelian structure.
Note that the finite value of our order parameter, given in lattice
units, indicates the occurrence of an AF condensate, 
$<A_\mu>=\Lambda O(g^{-1})$, which is
strong enough to modify the usual beta-function in the
UV scaling regime, \cite{magriv}.
\begin{figure}[htb]
\centering\epsfig{file=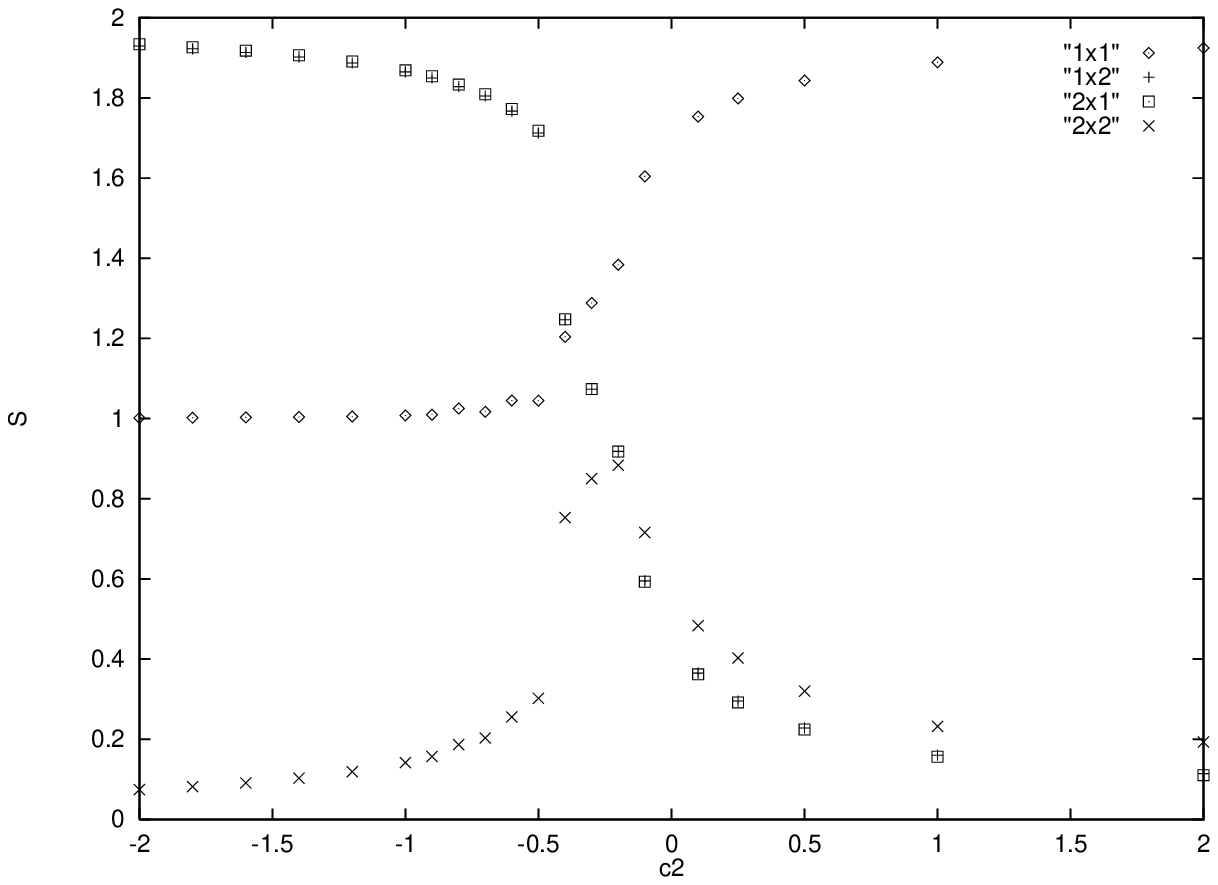,width=0.8%
\linewidth,bbllx=50,bblly=50,bburx=400,bbury=300}
\caption{Action density for $\beta=2.4$, $c_1=-1.0$ as a function
         of $c_2$. The third coupling was always adjusted so that
         $c_3=1/16-c_1/16-c_2/2$. Different symbols denote
         contributions from the $1\times1$, $1\times2$ and $2\times2$
         Wilson loops to the action.}
\label{fig:f6}
\end{figure}
\begin{figure}[htb]
\centering\epsfig{file=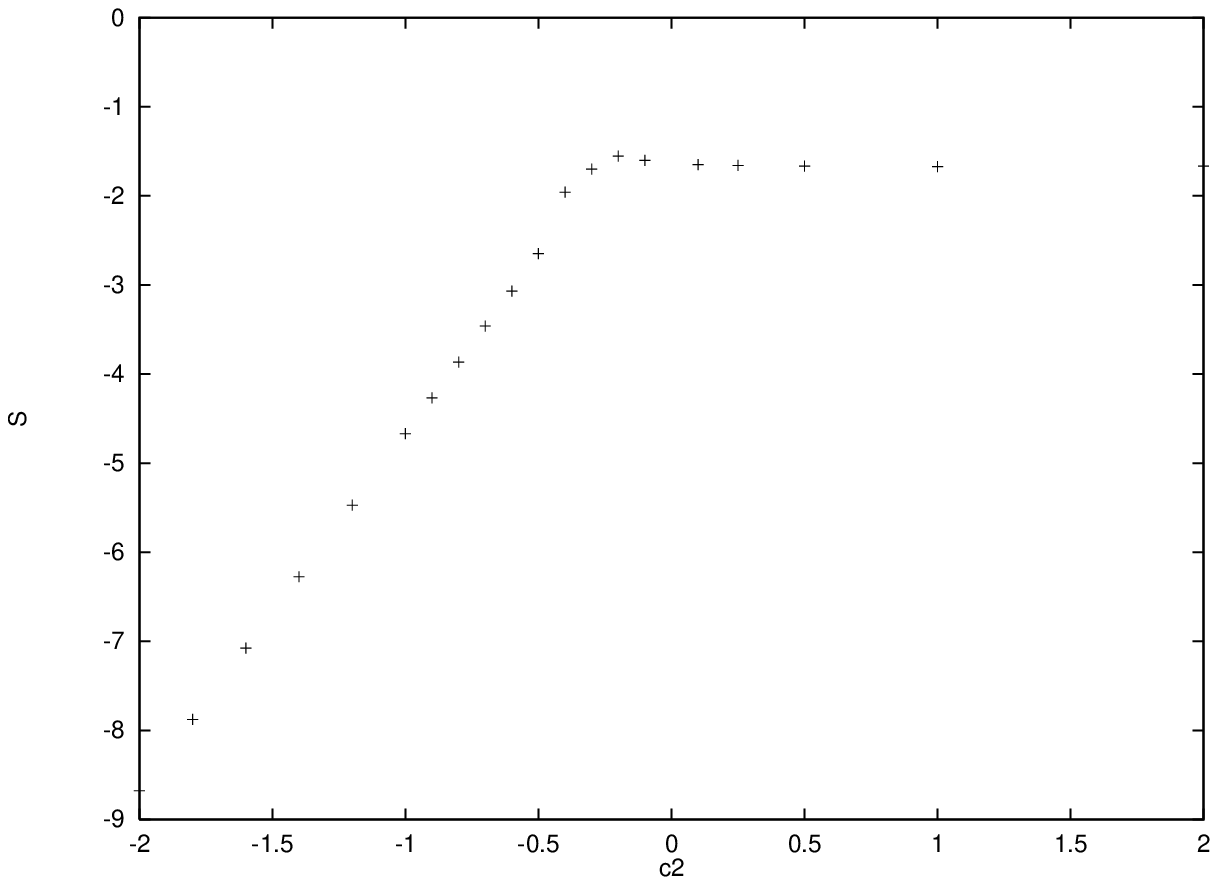,width=0.8%
\linewidth,bbllx=50,bblly=50,bburx=400,bbury=300}
\caption{Total action. According to eq.10 the trivial configuration,
         $U_\mu(x)=1$, has vanishing total action. Parameter values
         correspond to fig.6.}
\label{fig:f7}
\end{figure}
The total action $S$ corresponding to
fig.\ref{fig:f3} shown in fig.\ref{fig:f7}
indicates that $S$ is negative, i.e. it lies below the value
corresponding to the trivial vacuum, $A_\mu=0$,
due to the negative coupling constants. Fluctuations detected by
the operators, which multiply these coupling constants in the action
carry negative energy and condense. The qualitative behaviour of the
different terms of the action support the presence
of an AF ordering with period length $2a$.
In fact, the elongated, $1\times2$ loops change sign and the
$1\times1$ plaquette average settles halfway 
between the maximum and minimum as we move into the new phase.
At the same time the $2\times2$ loops
have a peak at the phase transition with the same
asymptotic value in both phases. This indicates that
the $2\times2$ loops,
which enclose two elementary cells, don't see the AF ordering.
In this case the total action became negative already before
the transition.

For sufficiently large FM coupling
we expect a series of phase transitions along orthogonal
directions, $\hat v_\alpha$ and $\hat w_\alpha$,
with increasing number of AF directions.
For a fixed value of the overall coupling $\beta=4/g^2>0$ and $\rho=0$
the system is in the FM phase.
Moving away from this point increasing $\rho$ at
fixed $\beta>0$ we expect to reach a series of AF phase boundaries
as the system becomes more and more frustrated.
It is possible to parameterize the locations of the $n$-th
phase transition as $c(\beta;~\rho_n(\beta,\phi),\phi)$.
As $n$ increases more and more planes should show AF ordering.
The topology of the phase boundaries is an interesting object for
further investigations.
It can consist of a system of concentric circles,
$\rho_n(\beta,\phi)<\rho_{n+1}(\beta,\phi)$,
or the phase boundary lines may touch each other, 
$\rho_n(\beta,\phi)=\rho_{n+1}(\beta,\phi)$ for certain $\phi$,
or they can have end points.
These end points might occur because the AF
order depends on whether the $1\times1$,
 $1\times2$ or $2\times2$ loop drives the instability. 

First numerical evidence for a complicated
topology of the phase boundary is shown
in figs.\ref{fig:f4} and \ref{fig:f5}.
The discontinuity of the order parameter in fig.\ref{fig:f3}
suggests a first order phase transition and the distribution of the
plaquette in fig.\ref{fig:f5} indicates that five planes became AF
at the same critical point. It is possible that
five phase boundary lines meet in this point.

\section{Weak Coupling Limit}
The phase transitions shown in the previous sections certainly change
the infrared behaviour of the vacuum for finite value
of the lattice spacing.
In this section we are concerned with the AF
order in the continuum limit, $a\to0$. A complete systematical
investigation of this limit is beyond the scope of this investigation
which was focused on two points:
\begin{itemize}
\item
the continuum limit is {\it different} in the AF
phase than in the usual one with $c_\alpha=0$ and
\item
this difference can be understood only by assuming the existence of relevant
operators different than $(F_{\mu\nu}^a)^2$. This is similar to the
case of the two dimensional Ising model where the the NNN coupling
with AF sign is relevant in the non-homogeneous vacuum.
\end{itemize}
 
Let us first suppose that the system has a critical point and
remains asymptotically free, i.e. the correlation length diverges
for $\beta\to\infty$. The question is if
the phase transitions found at finite $\beta$ persist in the
limit $\beta\to\infty$?

\begin{figure}[htb]
\hbox{\epsfig{file=F3.ps,width=0.5\linewidth}
      \epsfig{file=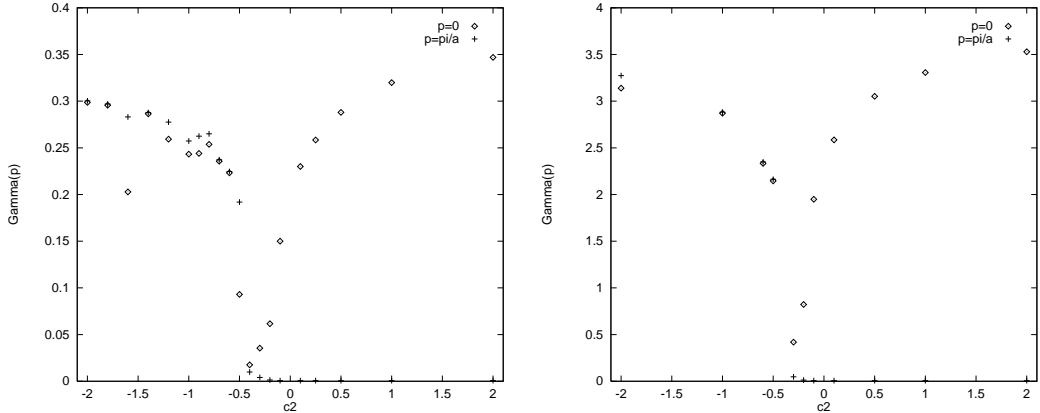,width=0.5\linewidth}}
\caption{Comparison of the normal ($p=0$) and staggered ($p=\pi/a$)
         order parameter as a function of $c_2$
         for $\beta=2.4$ (left) and $\beta=3.0$ (right),
         $c_1=-1$, and $c_3=1/16-c_1/16-c_2/2$.
         Note the different scale on the ordinates.}
\label{fig:f8}
\end{figure}

\noindent
Additional calculations of the order parameter
at higher $\beta$ show the same structure as
found earlier at $\beta=2.4$ (see fig. \ref{fig:f8}).
At larger coupling $\beta$ the correlation $\Gamma(p)$ becomes even stronger.
To confirm the stability of the transition
we performed further studies, not shown in the 
picture, at $\beta=2.5$ on larger lattices of size $12^4$. 
Again the structure did not change and we found only very small finite
size effects. Our numerical results
indicate that the anti-ferromagnetic phase transition 
is controlled by the couplings $c_1, c_2$ and $c_3$
almost independently from $\beta$.
As a consequence it is plausible that
the anti-ferromagnetic vacuum structure
persists the continuum limit.

\noindent
In the following it is instructive to look at the energy-entropy balance 
of SU(2) gauge theory with plaquette terms 
in the fundamental and adjoint representation \cite{bhancr}.
For sufficiently large positive coefficient of the
adjoint action, $\beta_A>\beta_{Acr}$, one finds a phase boundary
separating the weak from the strong coupling regime.
The transition is driven by a gas of fluxons.
The fluxon gas can be constructed by first setting
all link matrices to the identity and changing some of them to a
non-trivial center element of the gauge group \cite{toni}.
These configurations satisfy the lattice equations of motion for any action
made of Wilson loops. They become unstable for $\beta_A<\beta_{Acr}$
\cite{bhda}. 
Fluxons give rise to positive contributions to the action coming from the 
plaquette term in the fundamental representation. They are degenerate
with the vacuum, $U_\mu(x)=1$, as far as the adjoint plaquette is 
concerned. The phase transition occurs when these lattice artifacts
become suppressed as we approach the weak coupling regime.
As a result of this suppression the phase transition has no relevance
for the continuum theory. 

The lesson from this generic example is
that the local excitations have a chance to survive the
limit, $\beta\to\infty$, if their energy does not suppress them.
This is what happens in the AF phase because it occurs at negative value 
of the action density. Since the action density is bounded from below 
there must be AF solutions of the equations of motion at negative
action, i.e. below the trivial vacuum, $A_\mu=0$. This semi-classical
vacuum is highly non-trivial. The solutions of the lattice equations
of motion with minimal value of the action are not yet known.
It is useful to parameterize the
coupling constants by polar angles $(\theta,\phi)$ of the 
infinite sphere $S_2^c$ as $c(\beta,\tan\theta,\phi)$
in the limit $\beta\to\infty$.
Quantum fluctuations at the scale of the cut-off freeze out in this
limit leaving the short range structure of the semi-classical vacuum unchanged
because the modification can be expressed by a multiplicative constant
in the action. According to fig.\ref{fig:f7},
inhomogeneous classical solutions are energetically preferred against
homogeneous ones for certain values of the parameters $\theta$ and $\phi$.
Thus, the nontrivial, short-range order of the vacuum found
at intermediate values of $\beta$ survives the limit $\beta\to\infty$.
We expect that this change in the short range 
structure of the vacuum generates new phase transitions,
i.e.{} non-analytical
behaviour of densities as functions of the bare coupling constants.

According to the usual scenario there is only one relevant operator,
$F^2_{\mu\nu}$, in four dimensional gauge theory
and its coupling constant,
$g^2$, is asymptotically free. The remaining coupling constants in
eq.\ref{eq:extac} are irrelevant and may be kept constant as the cut-off
is removed. The actual choice of the irrelevant coupling constants only
influences the numerical value of the scale parameter, $\Lambda_{QCD}$.
Our main point is that this picture is incomplete. Indeed, it also
requires that the dynamics differs only by an overall scale on the FM
and AF side of the phase transition on $S_2^c$. Such an equivalence
between FM and AF phases is not observed even in simpler
models because the two condensates couple differently to the particle
modes. The difference between the phases should become even more 
pronounced when staggered quarks are introduced since they
correspond to an AF vacuum \cite{semenoff} by themselves.
Interactions between the excitations above the AF gauge vacuum
and the staggered fermions may even be used to split the unwanted
degeneracies and reduce the fermionic species doubling.

If we ask the question to what extent the usual scenario is modified
by the extended action given by eq.\ref{eq:extac}, the following cases
are possible:
\begin{enumerate}
\item {\it Renormalizable and Asymptotically Free:} If asymptotic
scaling prevails in the AF phase the continuum
limit corresponds to $\beta=\infty$. The coupling constants,
which are responsible for the AF condensate are relevant
because they modify the theory by more than an overall change of scale.
A higher dimensional operator may become relevant if its
anomalous dimension is sufficiently negative.  The lesson
from the semi-classical approximation of ref.~\cite{enzo}
is that the ratio of the coupling constants $c_2^2/c_4$ appears 
in the observables in addition to the usual polynomial terms, 
$c_2^mc_4^n$, arising from the perturbation expansion.
This ratio which can be large even if the $c_\alpha$ are small may
be the source of the large anomalous dimensions.
The AF short distance structure of the vacuum is calculable in the framework of 
a saddle point expansion because $\beta$ is large.
The long range order is lost due to the infrared
slavery as in the integer spin AF Heisenberg chain
\cite{heisenberg}. The renormalization of the theory reveals new ultraviolet
divergences due to the rapid oscillation in the semi-classical vacuum.

\item {\it Renormalizable and Non-asymptotically Free:}
Due to changes in the ultraviolet structure of the theory asymptotic
freedom may be lost.
The continuum limit is already reached at finite value of $g^2$
and strong quantum fluctuations may introduce lattice
defects and disorder.
The vacuum is highly non-trivial both in the UV and the IR region.

\item{\it Non-renormalizable:} The model may loose all
critical points in the 
coupling space and become non-renormalizable. It can be considered
as an effective theory up to a certain scale, the UV Landau pole. 

\end{enumerate}

When either of these possibilities is realized a
genuinely new phase of the
underlying gauge theory is identified. To find out,
which option is actually
realized goes beyond the scope of the present paper and 
clearly requires further work.

\section{Conclusion}
We investigated the
role of higher order derivatives in non-Abelian gauge theories.
It was found that the relevance or irrelevance of such terms 
is a highly non-trivial issue, which cannot be settled on the basis of 
simple perturbative arguments, because higher order derivatives
may enhance the contribution of configurations with characteristic
scale close to the cut-off. We found numerical evidence that
the semi-classical vacuum contains a condensate of modes close to the
cut-off in a certain region of the coupling constant space. 
In this region the field strength tensor shows oscillatory behaviour
and thereby displays AF, staggered ordering.
We put forward an argument that at least one of the possibilities,
the AF phase transition survives the removal of the cut-off
or the theory looses asymptotic freedom, is realized.

The renormalization of models with an AF vacuum is highly non-trivial
and cannot be covered by a simple extension of commonly used
perturbative methods developed for homogeneous or slowly varying 
background fields. We do not, at the present, possess a
procedure powerful enough to carry out the entire renormalization program.
Nevertheless, we believe that the Bloch-wave formalism provides
at least the framework in which this issue can ultimately be clarified.
Our present result should be considered only as an indication 
of the possibility of constructing a new type of continuum Quantum
Field Theory. A detailed analytical investigation is required
to reach a comprehensive understanding of the importance of the
anti-ferromagnetic vacuum structure.

The AF vacuum is generated by higher covariant derivative
operators in then action.
The relevance of operators depends on the global or long range
structure of the vacuum. The lattice gauge
theory defined by a modified plaquette action, 
\be
S_\kappa=\beta\sum_{x,\mu,\nu}\Theta\biggl(\kappa-
\Tr\left(1-\plaq\right)\biggr)~~
\Tr\left(1-\plaq\right)\nonumber
\label{eq:jocha}
\ee
as suggested in \cite{jochen} belongs to the same universality class
for any choice of the parameter $\kappa$ for $\beta>0$. 
This is certainly not true
in the presence of higher order derivatives in the AF
phase. In general, all higher order derivative terms in the action are 
irrelevant for the quantum fluctuations around the homogeneous
vacuum, $A_\mu=0$. They may become relevant when the saddle point 
is not translation invariant \cite{enzo,afisrg,afphrg}.

Higher order terms of the Lagrangian are supposed to arise from
the elimination of a heavy particle. The Peierls dimerisation
in a 1-dimensional electron-phonon system is an example where
vertices generated by a heavy fermion drive the effective
theory into an AF phase. Similar phenomena may occur in higher
dimensions as well \cite{brpo}. 

This remark leads us to our last point,
the striking similarity between 
the AF phase and solid state physics. 
In solid state physics, elimination of the ions generates
higher order derivatives in the effective theory of photons
and electrons. It is instructive to use our intuition from
Solid State Physics to imagine the dynamics of 
excitations above an AF vacuum. It has been already mentioned that the 
unitarity is lost for theories with higher order derivatives.
The effective theory of solids, involving
electrons, photons and phonons is highly non-unitary at high energies,
where ions appear as true degrees of freedom. Similarly, non-unitary
aspects of the gauge theory are relevant only at energies, where
AF ordering is destroyed by hard excitations.

Some well known phenomena are found in the AF phase.
Consider massless QED
with higher order terms in the photon action,
which make the semi-classical
photon vacuum AF. The electron spectrum develops
forbidden zones in the spectrum due to destructive Bragg reflections. 
One can introduce a chemical potential for the fermion number such
that the surface would be in the middle of a forbidden zone.
The resulting $\gamma^5$ symmetrical gap in the spectrum can be 
interpreted as a mass gap if the level density is non-vanishing at the
Fermi surface. In this way it is possible to have mass generation
in a vacuum, which seems homogeneous for infrared observations.
Another interesting effect may come from interactions between electrons
and small fluctuations of the photon field around the inhomogeneous
vacuum.
Fluctuations of $A_0$ create the Coulomb interaction between electrons.
By an appropriate fine tuning of the coupling constants of higher order 
terms in the action one can create an "acoustic branch", a massless
mode of transverse photons even in the absence of 
a gap in the one electron spectrum. Such unscreened, long range force 
may be attractive and play the role of phonons in generating
Cooper-pairs and super-conductivity in the vacuum.

\section{Note Added}
After completing the manuscript we noticed a preprint with similar
starting point but different goals \cite{shamir}.

\section{Acknowledgment}
We thank V. Branchina, D. Dyakonov, J. Jers\'ak and H. Mohrbach for
useful discussions.

\end{document}